\documentclass[twocolumn,a4paper]{article}

\usepackage{subfigure}
\usepackage{tikz}
\usetikzlibrary{shapes}
\usepackage{pgfplots}
\usepackage{pgfplotstable}
\usepackage{calc}
\usepackage{ifthen}

\begin{document}

\title{Teaching Software Engineering through Robotics}

\author{
Jiwon Shin \hspace*{5mm} Andrey Rusakov \hspace*{5mm} Bertrand Meyer \\
       Chair of Software Engineering\\
       Department of Computer Science\\
       ETH Zurich, Switzerland\\
       \{jiwon.shin, andrey.rusakov, bertrand.meyer\}@inf.ethz.ch
}

\maketitle
\begin{abstract}
This paper presents a newly-developed robotics programming course and reports the initial results of software engineering education in robotics context. Robotics programming, as a multidisciplinary course, puts equal emphasis on software engineering and robotics. It teaches students proper software engineering -- in particular, modularity and documentation -- by having them implement four core robotics algorithms for an educational robot. To evaluate the effect of software engineering education in robotics context, we analyze pre- and post-class survey data and the four assignments our students completed for the course. The analysis suggests that the students acquired an understanding of software engineering techniques and principles.

\end{abstract}

\section{Introduction}

Technological advancement has extended the importance of good software engineering far beyond the traditional computing devices and fields. Most university-level software engineering courses are, however, only for computer science students and do not emphasize the need for quality software in other fields. As a result, computer science students rarely gain hands-on experience with a real system, and students of other fields may only learn how to \emph{code} but not how to \emph{engineer} software. Robotics is one of these fields, where software engineering is an essential component, but the importance of software engineering education is overlooked.

From perception to control, robots fundamentally rely on computer programs to achieve desired behaviors. As robots advance and their complexity grows, robotics field's reliance on software libraries and thereby, its need for high-quality software augment. So far, however, roboticists have mainly focused on algorithmic development; standard software quality requirements such as modularity, correctness, and robustness have not generally been a top concern. Consequently, university-level robotics courses teach robotics algorithms but cover little to no software engineering topics.

In this paper, we present and evaluate a newly-developed, multidisciplinary course that teaches students software engineering and robotics. To our knowledge, our course, taught for the first time in the fall of 2013, is the first course to combine software engineering and robotics equally. The course teaches master's students in computer science, electrical engineering, and mechanical engineering how to bring good software engineering practices to robotics. Students apply software engineering techniques by implementing four core robotics algorithms for an educational robot. Through the course, students not only gain hands-on experience in a multidisciplinary environment but also learn to appreciate good software engineering techniques.

We evaluate the course to understand if student can learn software engineering when learning software engineering in robotics context. In particular, we investigate whether or not the students can
\begin{itemize}
\item learn the importance of software engineering principles, in particular, modularity and documentation
\item apply software engineering techniques in programming a real system, and
\item value good software engineering practice.
\end{itemize}
To answer these research questions, we conducted a pre- and post-class surveys and analyzed the four assignments students completed during the semester. The pre-class survey collected the students' educational background through multiple-choice questions. The post-survey had open-ended questions about various aspects of the course, and in this paper, we concentrate primarily on the two questions that are directly related to software engineering. The students' written responses to the questions are analyzed using thematic analysis~\cite{bib:braun-qrp-2006}. We analyze the four assignments using software quality metrics~\cite{bib:henderson-sellers-oom-1996} to evaluate the improvement of software quality quantitatively. Our analysis shows that students have gained an understanding and appreciation of good software engineering techniques.

The remainder of the paper is organized as follows. Section~\ref{sec:related} presents related work. The paper continues with a description of the course in Section~\ref{sec:course} and the methodology in Section~\ref{sec:methodology}. Section~\ref{sec:results} presents the results, and Section~\ref{sec:discussion} discusses them. Section~\ref{sec:threats} discusses possible threats to validity. The paper concludes with final remarks in Section~\ref{sec:conclusions}.

The proposed robotics programming course is a multidisciplinary course that combines software engineering and robotics. The importance of multidisciplinary engineering education was noticed over a decade ago~\cite{bib:lee-jc-1998}, and recently, Hotaling et al.~have shown the benefits of multidisciplinary engineering education in preparing undergraduate students for their future~\cite{bib:hotaling-jee-2012}. In particular, as a multidisciplinary field, robotics has gained attention as a medium to teach science and engineering concepts~\cite{bib:beer-acmc-1999}. Weinberg et al.~propose to use robotics to provide hands-on instruction in various fields of engineering and computer science education and overcomes the challenges of teaching this multidisciplinary course through multidisciplinary cooperation~\cite{bib:weinberg-eeac-2001}. Similarly, our course reduces the multidisciplinary difficulty by having multidisciplinary the teaching staff: of the two instructors and two teaching assistants, one instructor and one assistant have background in software engineering, and the other instructor and assistant in robotics.

Our course is a master's-level elective course that emphasizes proper software engineering and robotics algorithms. So far, however, most research on education of computer science and robotics fall into one of two categories: introductory computer science course that uses robotics as a medium to teach core computer science and/or programming concepts, or robotics courses with an emphasis on computer science that teach computational and algorithmic aspect of robotic software. In the former category of teaching introductory computer science through robotics, Fagin and Merkle show the negative effect of teaching computer science through robotics and attribute the lack of personal robots as the main reason for the failure~\cite{bib:fagin-sigcse-2003}. Since then, many others have used robotics to teach introductory computer science classes with more positive outcome~\cite{bib:summet-sigcse-2009, bib:markham-iticse-2010, bib:mcgill-tce-2012}. Like the successful introductory courses, our course also provides a complete robotic set to every student and allow the students to keep the hardware for the entire duration of the course. In addition, as an upper-level course, our course goes deeper into software engineering concepts such as design patterns, architecture, concurrency, and tools.

In the second category of robotics courses with computer science components, most courses focus on the computational/algorithmic aspect of robotics~\cite{bib:lalonde-icra-2006, bib:correll-toe-2013}. Popular robotics textbooks may teach specific topics of robotics~\cite{bib:thrun-pr-2005, bib:lavalle-pa-2006} or general robotics topics with some computational principles~\cite{bib:siegwart-amr-2011}, but they hardly mention of software engineering aspect of the algorithm implementation. Although Gustafson proposed using robotics to teach software engineering~\cite{bib:gustafson-fie-1998}, most robotics courses teach students \emph{what} to implement but overlook the question \emph{how} to implement them. 

To our knowledge, the only one other course, offered once in 2010 at GeorgiaTech, specifically targets to teach students how software engineering applies to robotics~\cite{bib:ser-2014}. While the general objectives of the course is similar to ours, the course content differs significantly. GeorgiaTech course focuses mostly on software engineering with robotics as a medium to teach software engineering concepts. The course's coverage on computational aspect of robotics is limited to robot control and navigation. Our course, on the other hand, makes equal emphasis on both software engineering and robotics. We cover software engineering and robotics topics equally, and the assignments require students to implement four core robotics algorithms: control and obstacle avoidance, localization, path planning, and object recognition. Students can therefore learn the core aspects of robotics while applying software engineering principles into practice.

We evaluate the course's ability to teach software engineering by combining qualitative analysis and quantitative analysis. Among various data collection techniques for field studies of software engineerings~\cite{bib:lethbridge-ese-2005}, we conduct questionnaires (surveys) in the beginning and at the end of the course and analyze the four assignments the students submitted during the course. The pre-survey had multiple choice questions and the post-survey had open-ended questions. To analyze the open-ended questions, we draw the inspiration from McCartney, Gokhale, and Smith~\cite{bib:mccartney-icer-2012} and apply thematic analysis~\cite{bib:braun-qrp-2006}. Unlike the authors, we strengthen the qualitative analysis by analyzing the student-generated code using software quality metrics~\cite{bib:henderson-sellers-oom-1996}.

\section{Course Description}
\label{sec:course}

Software engineering is a key component of modern systems, but teaching of software engineering has been limited to students of computer science. This section describes a new, multidisciplinary robotics programming course that teaches students  software engineering and robotics. The course provides computer science students an opportunity to apply their software engineering skills on a real system and teaches non-computer science students, in particular, students of robotics, proper software engineering. 

\subsection{Objectives}
\label{subsec:objectives}
The main objectives of the course are that students gain hands-on experience by programming a small robotic system with aspects of sensing, control, and planning, and gain knowledge of
\begin{itemize}
\item basic software engineering principles and methods,
\item the most common architectures in robotics,
\item coordination and synchronization methods, and
\item how software engineering applies to robotics.
\end{itemize}
As a multidisciplinary, master's-level, elective course, we intend our students, who come from computer science, electrical engineering, and mechanical engineering, to learn from one another and deepen their understanding of software engineering and robotics. To address the multidisciplinary need, the course is taught by a software engineer and a roboticist and assisted by graduate students in software engineering and robotics.

\subsection{Demography}
\label{subsec:demography}

The course is a master's-level elective course. The enrolment is open to any student of computer science, mechanical engineering, and electrical engineering, with some programming experience. Due to its resource intensive nature, the course is limited to 16 students. The first offering had a total of 12 students, 11 of who completed the course successfully. Of the 11 students, four were computer science students, six were mechanical engineering students, and one was a electrical engineering student. Two students were bachelor students and nine were master students.

The teaching staff comes from diverse background to reflect multidisciplinary nature of the course. In particular, the main lecturers consist of a software engineer and a roboticist and the teaching assistants are a graduate student in software engineering and a graduate student in mechanical engineering specializing. Diversity in background enables the teaching staff to better understand difficulties students would face in the course and address their needs.

\subsection{Content}
\label{subsec:content}
Lecture topics for the course ensure balanced exposure to both software engineering and robotics. Software engineering lectures include concurrency, design patterns, modern software engineering tools, and software architecture in robotics. Robotics lectures are robot control and obstacle avoidance, localization and mapping, path planning, and object recognition. In addition, there is a lecture on Robot Operating System (ROS)~\cite{bib:quigley-iros-2009}, a popular middleware in robotics, and our robotics framework.

\subsection{Assignments}
\label{subsec:grading}
The course contains five assignments: an ungraded assignment for setting up the environment and four graded assignments, each implementing a core robotics algorithm. The initial ungraded assignment gives students a chance to get familiar with the hardware and software environment. The four graded assignments require students to implement algorithms for control and obstacle avoidance, localization, path planning, and object recognition. The first two graded assignments are individual assignments to encourage every student to learn the basics. The last two graded assignments are completed as a team of two students to allow the students to work on more extensive tasks.

Object-oriented programming and concurrency are the key programming paradigms for the class. As the object-oriented programming language, the course uses Eiffel and C++. The first assignment is done entirely in Eiffel, and the remaining three assignments have both Eiffel and C++ components. Eiffel with its simple concurrency extension, SCOOP, enables students to learn and program concurrent software easily~\cite{bib:nanz-esem-2011}. C++, as a popular programming language in robotics and one of the main languages of ROS~\cite{bib:quigley-iros-2009}, enables students get familiar with a common robotics programming environment.

The course emphasizes correct implementation of robotics algorithms as well as proper software engineering. To emphasize the importance of functionality and quality, the assignments are graded equally on the demonstration of the implemented algorithm (50\%) and the software quality (50\%). The students submit their software via their SVN repository and demonstrate in class how their implementation works on their robot. The grading scheme for the software quality portion is as follows:
\begin{itemize}
\item Choice of abstraction and relations (30\%)
\item Correctness of implementation (40\%)
\item Extendibility and reusability (20\%)
\item Comments and documentation (10\%)
\end{itemize}

To ensure continuous learning throughout the course, students receive two types of feedback after each assignment: in-class feedback and individual feedback. After each assignment, students learn in class about common mistakes and ways to avoid them in the future. In addition, we conduct individual feedback sessions. During the 15-minute long individual feedback session, students receive feedback on their particular software's  correctness and quality. Immediate feedback enables students to incorporate newly-acquired knowledge into subsequent assignments and improve their software practice over time.

\section{Methodology}
\label{sec:methodology}

This pilot study investigates if students can learn software engineering principles and techniques in a robotics programming class. In particular, we are interested in investigating if our students can learn the importance of software modularity and documentation. To answers these research questions, we conduct a pre-class and a post-class survey and analyze the four graded assignments the students completed for the course. The data are collected from a single-offering of the course, held in Fall 2013. This section describes the study design, data collection, and analysis method.

\subsection{Surveys}

We conducted pre-class survey and post-class surveys to evaluate effectiveness of teaching software engineering in the robotics programming course. The pre-class survey collected the students' educational background in software engineering, robotics/control theory, and computer vision as multiple-choice questions. The post-class survey had open-ended questions about software engineering, robotics, and overall experience. Neither survey was anonymous, and no incentive was given for completing the survey.

In this pilot study, we are particularly interested in investigating if the students understand the value of proper software engineering. We therefore analyze a subset of the survey responses that are related to software engineering. In the pre-class survey, we analyze the students' background in software engineering, i.e., programming, object-oriented programming, and concurrent programming experience and knowledge of software engineering concepts and tools. In the post-class survey, we analyze the responses to the following two questions:
\begin{itemize}
\item Which software engineering concepts and tools did you use in this course? Any examples?
\item Has your approach to the assignment changed over the course of the semester? Do you think about your software's architecture before you begin the implementation?
\end{itemize}

We collected data from all the students who completed the course successfully. The course began with its full capacity of 16 students, but after the first two weeks of the semester, the number of students was down to 12. Of the 12 students, 11 completed the course successfully. The pre-class survey was conducted in class during the second week of the semester, and of the 11 successful students, 10 filled it out. The post-class survey was sent to the students as e-mail at the end of the semester. The students had a week to fill out and return the survey. With a friendly reminder, all 11 students returned the post-class survey.

We analyze the responses to the multiple-choice questions of the pre-class survey by counting the number of occurrences. To understand the responses to the open-ended questions of the post-class survey, we employ thematic analysis~\cite{bib:braun-qrp-2006} to capture the main themes in the responses. We follow the procedure as suggested by Braun and Clarke and identify the themes by familiarizing ourselves with the responses, coding text fragments, grouping recurring fragments, then combining related groups into themes. To represent the underlying data correctly, we ignore rarely-occurring text fragments and verify that themes and sub-themes capture frequently occurring text fragments.

\subsection{Assignments}

The main objective of analyzing the assignment is to understand how the students improved quality of their software over the semester. To evaluate, we analyze the four graded assignments the students completed and submitted during the semester. Each assignment required the students to implement a core robotics algorithm, and the students received their grades based on in-class demonstration of the implemented algorithm and quality of the implemented software. In this pilot study, we investigate the change in the software quality over the four graded assignments.

The students submitted their assignments via their Subversion (SVN) repository. The first two assignments were individual assignments while the last two assignments were done in a team of two. Students had three weeks to complete the first assignment, four weeks for the second assignment, two weeks for the third assignment, and two and a half weeks for the final assignment. Varying duration reflects difference in difficulty among the assignments.

To evaluate the improvement of software quality over time, we use software metrics. Software metrics enable quantitative analysis of software. Many metrics have been suggested for measuring software quality, but no consensus exists~\cite{bib:henderson-sellers-oom-1996}. Of many metrics, we pick a subset of the commonly-used metrics. In addition, we select some metrics to capture common mistakes of our students. Metrics used for analysis are percentage of comments, number of arguments per routine, percentage of routines with hard-coded values, lack of parametrization, and number of SVN commits. The relation between the metrics and software quality is as follows:
\begin{itemize}
\item Although some argue that having comments is a sign of bad software, we believe that comments are essential to reusable software. We consider software with a higher percentage of comments more reusable, but software with more lines of comments than code less reusable.
\item Number of arguments per routine is also related to reusability. We consider routines with fewer arguments more reusable.
\item Hard-coded values reduce reusability and extendibility. We consider software with fewer or no hard-coded values to be more reusable and extendible.
\item Ability to set parameters outside of a routine or a class makes software reusable and extendible. We consider software without parametrization less reusable and extendible.
\item SVN commits reflect software engineering practice, and we consider small, frequent commits to be better than large, infrequent commits.
\end{itemize}

The set of metrics we select for this analysis is neither complete nor the most representative metrics for software quality. We select these metrics for their measurability and clear link to software quality. We believe that software that perform better on these metrics is of higher quality than software that do not perform as well on them, however marginal the improvement may be.

\section{Results}
\label{sec:results}

This section presents the results of the pre- and post-class surveys and the analysis of the assignments.

\subsection{Pre-class Survey}

Of the 11 students who completed the course successfully, only 10 students filled out the pre-survey. Eight were master's students, and two were upper-bachelor's students. Six students were mechanical engineering students, three were computer science students, and one was an electrical engineering student. Only one of the eight master's students had completed his bachelor's degree at ETH Zurich. The student whose data are not included in the pre-class survey results was a master's-level computer science student with a bachelor's degree from ETH Zurich. 

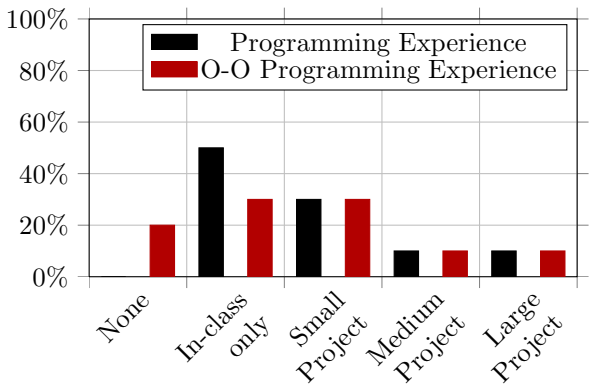
\begin{figure}[h!]
	\centering

	\begin{tikzpicture}
	\begin{axis}[
	width=8cm, height=5cm,
	 scaled ticks=false, tick label style={/pgf/number format/fixed},
	 xbar, 
	 xmin = 1, xmax = 6,
	 ymin = 0, ymax = 100,
	 ytick = {0, 20, 40, 60, 80, 100},
	 yticklabels = {0\%, 20\%, 40\%, 60\%, 80\%, 100\%},
	 xtick=data,
	 xticklabel style = {inner sep=0pt,anchor=north east,rotate=45,text width=35pt,align=right},
	 xticklabels = {None, In-class only, Small Project, Medium Project, Large Project},
	 ybar interval=0.5,
	 bar width=3pt,
	 grid = major,
	 legend style={cells={anchor=center, fill}, nodes={inner sep=1, below=-1.1ex}}, area legend
	 ]
	  \addplot [color = black, fill = black]	
	    coordinates{(1,0)
	                (2,50)
	                (3,30)
	                (4,10) 
	                (5,10)
	                (6,0)};
	  \addplot [color = red!70!black, fill = red!70!black]
	    coordinates{(1,20)
	                (2,30)
	                (3,30)
	                (4,10) 
	                (5,10)
	                (6,0)};
	 \addlegendentry{Programming Experience}
	 \addlegendentry{O-O Programming Experience}
	 \end{axis}
	\end{tikzpicture}

	\vspace{-4mm}
	\caption{Prior programming experience (black) and object-oriented programming experience (red) of the students. N=10}
	\vspace{-4mm}
	\label{fig:experience}
\end{figure}

Fig.~\ref{fig:experience} shows the distribution of programming and object-oriented programming experience our students had prior to starting the course. The course require some programming experience, and 80\% of the students indicated that their programming experience is limited to class assignments or small projects consisting of less than 10'000 lines of code. In terms of prior object-oriented programming experience, 20\% of the students had none, and 60\% had little experience. This meant that most students had learn object-oriented paradigm before they could complete the assignments.

\begin{figure}[h!]
	\centering
	
	\begin{tikzpicture}
	\begin{axis}[
	width=8cm, height=3cm,
	 scaled ticks=false, tick label style={/pgf/number format/fixed},
	 xbar, 
	 xmin = 1, xmax = 6,
	 ymin = 0, ymax = 60,
	 ytick = {0, 20, 40, 60},
	 yticklabels = {0\%, 20\%, 40\%, 60\%, 80\%, 100\%},
	 xticklabel style = {inner sep=0pt,anchor=north east,rotate=45,text width=40pt,align=right},
	 xticklabels = {None, Message passing, Threading, Monitor, Mutex/ Semaphore},
	 ybar interval=0.4,
	 bar width=2pt,
	 grid = major
	 ]
	  \addplot [color = red!70!black, fill = red!70!black]	
	    coordinates{(1,50)
	                (2,20)
	                (3,50)
	                (4,30) 
	                (5,40)
	                (6,0)};
	 \end{axis}
	\end{tikzpicture}

	\vspace{-4mm}
	\caption{Experience in concurrency. N=10}
	\vspace{-2mm}
	\label{fig:concurrency}
\end{figure}
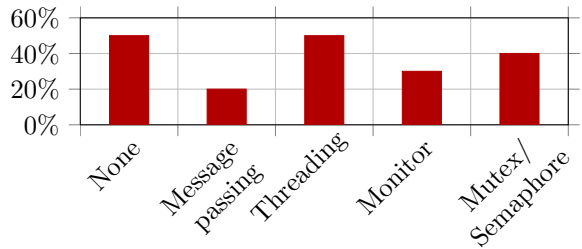

In terms of concurrency (Fig.~\ref{fig:concurrency}), half of the students had no experience while the other half had some. Of the people with some concurrent programming experience, everyone had worked with threads and most had used mutexes and/or semaphores. Every computer science student had some concurrent programming experience while most mechanical engineering student had none.

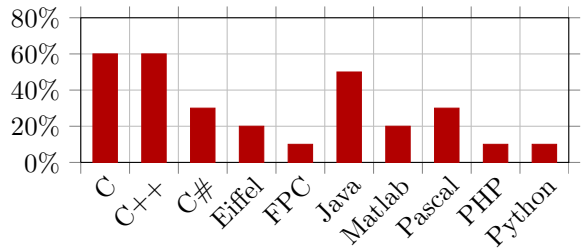
\begin{figure}[h!]
	\centering
	
	\begin{tikzpicture}
	\begin{axis}[
	width=8cm, height=3.5cm,
	 scaled ticks=false, tick label style={/pgf/number format/fixed},
	 xbar, 
	 xmin = 1, xmax = 11,
	 ymin = 0, ymax = 80,
	 ytick = {0, 20, 40, 60, 80},
	 yticklabels = {0\%, 20\%, 40\%, 60\%, 80\%, 100\%},
	 xtick = data,
	 xticklabel style = {inner sep=0pt,anchor=north east,rotate=45},
	 xticklabels = {C, C++, C\#, Eiffel, FPC, Java, Matlab, Pascal, PHP, Python},
	 ybar interval=0.5,
	 bar width=3pt,
	 grid = major
	 ]
	  \addplot [color = red!70!black, fill = red!70!black]	
	    coordinates{(1,60)
	                (2,60)
	                (3,30)
	                (4,20) 
	                (5,10)
	                (6,50)
	                (7,20)
	                (8,30)
	                (9,10)
	                (10,10)
	                (11,0)};
	 \end{axis}
	\end{tikzpicture}
	\vspace{-4mm}
	\caption{Familiar programming languages. N=10}
	\vspace{-2mm}
	\label{fig:languages}
\end{figure}

The students had experience with 10 different programming languages (Fig.~\ref{fig:languages}). C, C++, and Java were the most used. More than half of the students had programmed at least one of the aforementioned three languages. On average, computer science students knew 4.3 languages while other students knew 2.4 languages.

\begin{figure}[h!]
	\centering
	
	\begin{tikzpicture}
	\begin{axis}[
	width=8cm, height=3cm,
	 scaled ticks=false, tick label style={/pgf/number format/fixed},
	 xbar, 
	 xmin = 1, xmax = 6,
	 ymin = 0, ymax = 60,
	 ytick = {0, 20, 40, 60},
	 yticklabels = {0\%, 20\%, 40\%, 60\%, 80\%, 100\%},
	 xticklabel style = {inner sep=0pt,anchor=north east,rotate=45,text width=70pt,align=right},
	 xticklabels = {None, Programming Paradigms, Design Patterns, Algorithms/ Data Structure, Testing/ Verification},
	 ybar interval=0.4,
	 bar width=2pt,
	 grid = major
	 ]
	  \addplot [color = red!70!black, fill = red!70!black]	
	    coordinates{(1,12.5)
	                (2,37.5)
	                (3,50)
	                (4,50) 
	                (5,25)
	                (6,0)};
	 \end{axis}
	\end{tikzpicture}

	\vspace{-4mm}
	\caption{Software engineering concepts. N=8}
	\vspace{-5mm}
	\label{fig:concepts}
\end{figure}
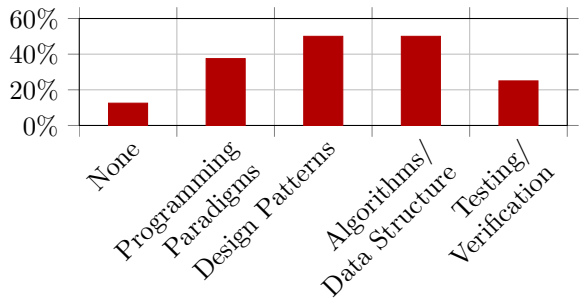

To the question on their exposure to software engineering concepts at a university, seven students indicated having learned one or more concepts (Fig.~\ref{fig:concepts}). One student answered not having learned any of the concepts, and two did not answer the question. On average, computer science students had been exposed to 3.5 concepts while other students were familiar with 1.2 concepts.

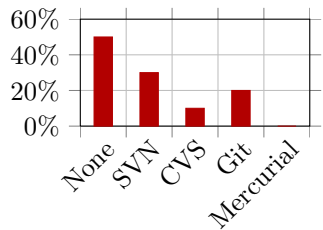
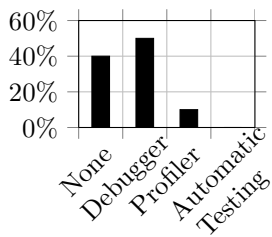
\begin{figure}[h!]
	\centering
    \subfigure[Configuration management]
    {	
	\begin{tikzpicture}
	\begin{axis}[
	width=4.6cm, height=3cm,
	 scaled ticks=false, tick label style={/pgf/number format/fixed},
	 xbar, 
	 xmin = 1, xmax = 6,
	 ymin = 0, ymax = 60,
	 ytick = {0, 20, 40, 60},
	 yticklabels = {0\%, 20\%, 40\%, 60\%, 80\%, 100\%},
	 xtick = data,
	 xticklabel style = {inner sep=0pt,anchor=north east,rotate=45,text width=50pt,align=right},
	 xticklabels = {None, SVN, CVS, Git, Mercurial},
	 ybar interval=0.4,
	 bar width=2pt,
	 grid = major
	 ]
	  \addplot [color = red!70!black, fill = red!70!black]	
	    coordinates{(1,50)
	                (2,30)
	                (3,10)
	                (4,20) 
	                (5,0)
	                (6,0)};
	 \end{axis}
	\end{tikzpicture}
	\label{subfig:config}
	}
	\subfigure[Software development]
    {
	\begin{tikzpicture}
	\begin{axis}[
	width=3.9cm, height=3cm,
	 scaled ticks=false, tick label style={/pgf/number format/fixed},
	 xbar, 
	 xmin = 1, xmax = 5,
	 ymin = 0, ymax = 60,
	 ytick = {0, 20, 40, 60},
	 yticklabels = {0\%, 20\%, 40\%, 60\%, 80\%, 100\%},
	 xticklabel style = {inner sep=1pt,anchor=north east,rotate=45,text width=35pt,align=right},
	 xticklabels = {None, Debugger, Profiler, Automatic Testing},
	 ybar interval=0.4,
	 bar width=2pt,
	 grid = major
	 ]
	  \addplot [color = black, fill = black]	
	    coordinates{(1,40)
	                (2,50)
	                (3,10)
	                (4,00)
	                (5,0)};
	 \end{axis}
	\end{tikzpicture}
	\label{subfig:setools}
	}
	\vspace{-4mm}
	\caption{Tools the students have used. N=10}
	\vspace{-2mm}
	\label{fig:tools}
\end{figure}

Fig.~\ref{fig:tools} shows the students' experience with tools. Half of the students had never used any configuration management tools, and others had used SVN or Git. One student indicated that he has used both SVN and CVS. In terms of software development, four students indicated that they have not used any software development tools. Five students indicated that they have used a debugger, and one student said that he has used both a debugger and a profiler. No one had any prior experience using an automatic testing tool.

\subsection{Post-class Survey}

We conducted the post-class survey to understand what the students had learned in the course. All eleven students completed the survey. We analyze their responses to two questions: concepts and tools they used, and change in their approach to software engineering. Their written responses are grouped into themes using thematic analysis~\cite{bib:braun-qrp-2006}.

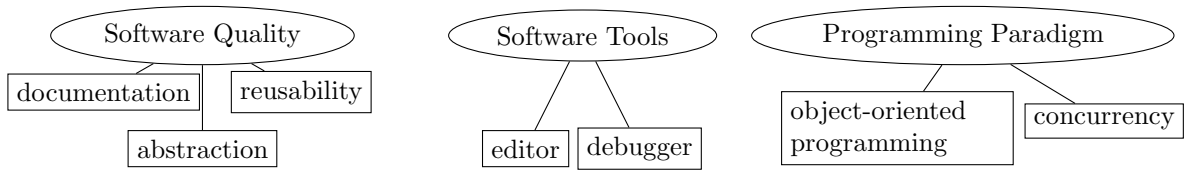
\begin{figure*}[t!]
	\centering

	\begin{tikzpicture}
		\node[ellipse,draw] at (0,0) {Software Quality}
			child[grow=-30] {node[rectangle,draw] {reusability}}
			child[grow=-90] {node[rectangle,draw] {abstraction}}
			child[grow=-150] {node[rectangle,draw] {documentation}};
		\node[ellipse,draw] at (5,0) {Software Tools}
			child {node[rectangle,draw] {editor}}
			child {node[rectangle,draw] {debugger}};
		\node[ellipse,draw] at (10,0) {Programming Paradigm}
			child[grow=-125] {node[rectangle,draw,text width=80pt] {object-oriented programming}}
			child[grow=-25] {node[rectangle,draw] at (0.5,-0.5) {concurrency}};
	\end{tikzpicture}
	\vspace{-3mm}
	
	\caption{Concepts and tools the students used in the class. We identified three main themes (ellipse) and several sub-themes (rectangle).}
	\vspace{-5mm}
	\label{fig:postsurvey}
\end{figure*}

Fig.~\ref{fig:postsurvey} shows the themes that we identified from the students' responses on concepts and tools they used. The three main themes are software quality, software tools, and programming paradigm. In software quality, the students mainly used reusability, abstraction, and documentation. In addition, code conventions, refactoring, extendibility, and readability got mentioned by one or two students. The students used editor and debugger as software tools. In programming paradigm, the students indicated using object-oriented programming and concurrency.

\begin{figure}
	\centering

	\begin{tikzpicture}
		\node[ellipse,draw,text width=60pt,align=center] at (0,0) {Software\\ Development}
			child[grow=-30] {node[rectangle,draw,text width=50pt] at (0,-0.5){software\\ architecture}}
			child[grow=-90] {node[rectangle,draw] at (0,-0.6){software quality}}
			child[grow=-150] {node[rectangle,draw,text width=50pt] at (0,-0.5) {incremental\\ development}};
		\node[ellipse,draw,text width=60pt,align=center] at (4,0) {Project\\ Management}
			child[grow=-30] {node[rectangle,draw,text width=50pt] at (0,-1) {time\\ management}}
			child[grow=-120] {node[rectangle,draw,text width=35pt] at (0,-0.5) {setting priority}};		
	\end{tikzpicture}

	\vspace{-3mm}
	\caption{Change in software engineering practice with main themes (ellipse) and sub-themes (rectangle).}
	\vspace{-4mm}
	\label{fig:se}
\end{figure}
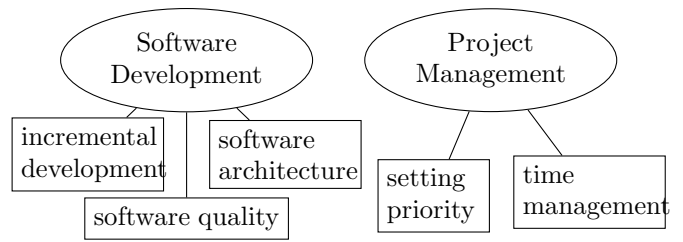

To the question about the change in their approach to the assignments over the course of the semester, 10 out of 11 students responded that they have changed their approach. As shown in Fig.~\ref{fig:se}, we identified two themes: software development and project management. The students improved their software development skills by thinking about software architecture, software quality, and/or applying incremental development. Many also improved project management skills in terms of time management and/or setting priority. Only one student, a master's student in computer science, indicated that there was no fundamental change to his work style as he ``always plan(s) the interface first''.

\subsection{Assignments}

\begin{figure}[h!]
	\centering
    \subfigure[Eiffel]
    {	
	\begin{tikzpicture}
	\begin{axis}[
	width=4.7cm, height=3.5cm,
	 scaled ticks=false, tick label style={/pgf/number format/fixed},
	 xbar, 
	 xmin = 1, xmax = 5,
	 ymin = 0, ymax = 25,
	 xtick = data,
	 xticklabel style = {inner sep=0pt,anchor=north east,rotate=45},
	 xticklabels = {A1, A2, A3, A4},
	 yticklabels = {0\%, 10\%, 20\%},
	 ybar interval=0.4,
	 bar width=2pt,
	 grid = major
	 ]
	  \addplot [color = red!70!black!70!white, fill = red!70!black!70!white]
	  [ error bars/.cd, y dir=both, y explicit,
	    error bar style={line width=1.5pt, red!40!black, xshift=4mm},
	    error mark options = {rotate = 90, line width=1pt, mark size = 3pt, red!40!black} ]	
	    coordinates{(1,6.0) +- (0,3.9)
	                (2,0.5) +- (0,0.6)
	                (3,0.6) +- (0,0.9)
	                (4,0.5) +- (0,0.7)
	                (5,0)};
	 \end{axis}
	\end{tikzpicture}
	\label{subfig:hardcodedeiffel}
	}
	\subfigure[C++]
    {
	\begin{tikzpicture}
	\begin{axis}[
	width=4cm, height=3.5cm,
	 scaled ticks=false, tick label style={/pgf/number format/fixed},
	 xbar, 
	 xmin = 1, xmax = 4,
	 ymin = 0, ymax = 25,
	 xtick = data,
	 xticklabel style = {inner sep=0pt,anchor=north east,rotate=45},
	 xticklabels = {A2, A3, A4},
	 yticklabels = {0\%, 10\%, 20\%},
	 ybar interval=0.4,
	 bar width=2pt,
	 grid = major
	 ]
	  \addplot [color = black!50, fill = black!50]
	  [ error bars/.cd, y dir=both, y explicit,
  	    error bar style={line width=1.5pt, black, xshift=4mm},
	    error mark options = {rotate = 90, line width=1pt, mark size = 3pt, black} ]	
	    coordinates{(1,12.4) +- (0,6.6)
	                (2,6.3) +- (0,3.0)
	                (3,9.3) +- (0,5.0)
	                (4,0)};
	 \end{axis}
	\end{tikzpicture}
	\label{subfig:hardcodedc++}
	}
	\vspace{-4mm}
	\caption{\% of routines with hard-coded values.}
	\vspace{-4mm}
	\label{fig:hardcoded}
\end{figure}
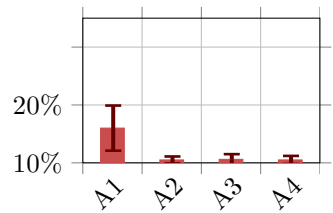
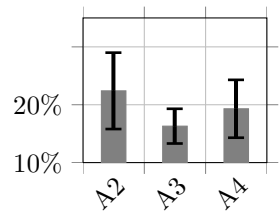

We analyze the four assignments the students completed during the semester to investigate if the students learned to apply good software engineering techniques. In particular, as our course emphasized software modularity and documentation, we evaluate the assignments using software quality metrics that are directly related to modularity and documentation. In addition, in grading the assignments, we identified some common mistakes that the students made:
\begin{itemize}
\item \emph{Single responsibility principle} violation. We observed the tendency of having a single big class with multiple unrelated functionalities, instead of separating them into several independent classes. This results in a seriously restriction of software reusability.
\item \emph{Unnecessary dependencies} between classes. Student-generated software often contained groundless relations between classes. These unnecessary dependencies limit further extension and reuse of the software.
\item Lack of \emph{parametrization}. We observed that for the first assignment, majority of the students implemented algorithms without taking extendibility into account and did not provide a way to set parameters outside of a routine or a class. This flaw not only restricts further reusability and extendability of the system but also makes testing and debugging more complicated.
\item \emph{Hard-coded} variables. Another common mistake, which directly affects both readability and reusability, was duplication of the same values in the code. The corresponding recommendation can be formulated as avoiding so-called "hard-coded" values and "magic numbers" by introducing variables and using language support for \textit{constants}. Fig.~\ref{fig:hardcoded} shows how the number of this mistakes have decreased at the end of the course.
\end{itemize}
Based on the observations, we add two more metrics to our evaluation. The final set of metrics are percentage of comments, number of arguments per routine, percentage of routines with hard-coded values, lack of parametrization, and number of SVN commits.

\begin{figure}[h!]
	\centering
    \subfigure[Eiffel]
    {	
	\begin{tikzpicture}
	\begin{axis}[
	width=4.7cm, height=3.5cm,
	 scaled ticks=false, tick label style={/pgf/number format/fixed},
	 xbar, 
	 xmin = 1, xmax = 5,
	 ymin = 0, ymax = 25,
	 xtick = data,
	 xticklabel style = {inner sep=0pt,anchor=north east,rotate=45},
	 xticklabels = {A1, A2, A3, A4},
	 yticklabels = {0\%, 10\%, 20\%},
	 ybar interval=0.4,
	 bar width=2pt,
	 grid = major
	 ]
	  \addplot [color = red!70!black!70!white, fill = red!70!black!70!white]
	  [ error bars/.cd, y dir=both, y explicit,
	    error bar style={line width=1.5pt, red!40!black, xshift=4mm},
	    error mark options = {rotate = 90, line width=1pt, mark size = 3pt, red!40!black} ]	
	    coordinates{(1,11.9) +- (0,2.8)
	                (2,9.6) +- (0,0.9)
	                (3,10.4) +- (0,1.3)
	                (4,10.6) +- (0,0.9)
	                (5,0)};
	 \end{axis}
	\end{tikzpicture}
	\label{subfig:commentseiffel}
	}
	\subfigure[C++]
    {
	\begin{tikzpicture}
	\begin{axis}[
	width=4cm, height=3.5cm,
	 scaled ticks=false, tick label style={/pgf/number format/fixed},
	 xbar, 
	 xmin = 1, xmax = 4,
	 ymin = 0, ymax = 25,
	 xtick = data,
	 xticklabel style = {inner sep=0pt,anchor=north east,rotate=45},
	 xticklabels = {A2, A3, A4},
	 yticklabels = {0\%, 10\%, 20\%}, 
	 ybar interval=0.4,
	 bar width=2pt,
	 grid = major
	 ]
	  \addplot [color = black!50, fill = black!50]
	  [ error bars/.cd, y dir=both, y explicit,
  	    error bar style={line width=1.5pt, black, xshift=4mm},
	    error mark options = {rotate = 90, line width=1pt, mark size = 3pt, black} ]	
	    coordinates{(1,16.7) +- (0,5.3)
	                (2,18.3) +- (0,4.4)
	                (3,18.4) +- (0,3.2)
	                (4,0)};
	 \end{axis}
	\end{tikzpicture}
	\label{subfig:commentsc++}
	}
	\vspace{-4mm}
	\caption{\% of comments.}
	\vspace{-5mm}
	\label{fig:comments}
\end{figure}
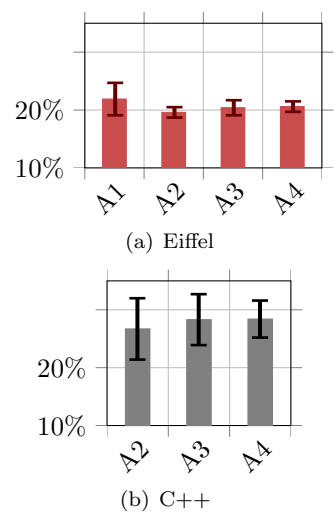

Comments improve understandability of code, which leads to increased reusability. As a metric for measuring comments, we compute the percentage of comments. The percentage of comments is the number of commented lines divided by the number of lines of code. This measure assumes that the code being analyzed is clean, but initially, many students submitted code without thorough clean-up. For accurate measurement, we ignored the commented out code blocks and only count the real comment lines.

Fig.~\ref{fig:comments} shows the percentage of comments for the four assignments. The students commented on average once every six to ten lines of code. The percentage of comments was higher for the first assignment than the other assignments in Eiffel, which may be due to the students reusing and modifying the example code we provided. The percentage of comments from the second assignment to the fourth assignment increased 1\% in Eiffel and 1.7\% in C++. 

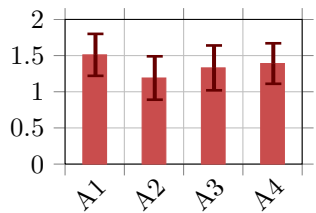
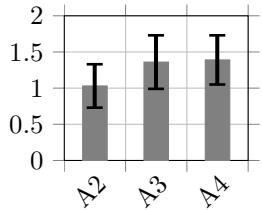
\begin{figure}[h!]
	\centering
    \subfigure[Eiffel]
    {	
	\begin{tikzpicture}
	\begin{axis}[
	width=4.7cm, height=3.5cm,
	 scaled ticks=false, tick label style={/pgf/number format/fixed},
	 xbar, 
	 xmin = 1, xmax = 5,
	 ymin = 0, ymax = 2,
	 xtick = data,
	 xticklabel style = {inner sep=0pt,anchor=north east,rotate=45},
	 xticklabels = {A1, A2, A3, A4},
	 ybar interval=0.4,
	 bar width=2pt,
	 grid = major
	 ]
	  \addplot [color = red!70!black!70!white, fill = red!70!black!70!white]
	  [ error bars/.cd, y dir=both, y explicit,
	    error bar style={line width=1.5pt, red!40!black, xshift=4mm},
	    error mark options = {rotate = 90, line width=1pt, mark size = 3pt, red!40!black} ]	
	    coordinates{(1,1.51) +- (0,0.29)
	                (2,1.19) +- (0,0.3)
	                (3,1.33) +- (0,0.31)
	                (4,1.39) +- (0,0.28)
	                (5,0)};
	 \end{axis}
	\end{tikzpicture}
	\label{subfig:argeiffel}
	}
	\subfigure[C++]
    {
	\begin{tikzpicture}
	\begin{axis}[
	width=4cm, height=3.5cm,
	 scaled ticks=false, tick label style={/pgf/number format/fixed},
	 xbar, 
	 xmin = 1, xmax = 4,
	 ymin = 0, ymax = 2,
	 xtick = data,
	 xticklabel style = {inner sep=0pt,anchor=north east,rotate=45},
	 xticklabels = {A2, A3, A4},
	 ybar interval=0.4,
	 bar width=2pt,
	 grid = major
	 ]
	  \addplot [color = black!50, fill = black!50]
	  [ error bars/.cd, y dir=both, y explicit,
  	    error bar style={line width=1.5pt, black, xshift=4mm},
	    error mark options = {rotate = 90, line width=1pt, mark size = 3pt, black} ]	
	    coordinates{(1,1.03) +- (0,0.3)
	                (2,1.36) +- (0,0.37)
	                (3,1.39) +- (0,0.34)
	                (4,0)};
	 \end{axis}
	\end{tikzpicture}
	\label{subfig:argc++}
	}
	\vspace{-4mm}
	\caption{Arguments per routine.}
	\vspace{-2mm}
	\label{fig:arguments}
\end{figure}

Fig.~\ref{fig:arguments} shows the average number of arguments for each assignment. Because every assignment required the students to implement a different algorithm, a direct comparison of the average number of arguments between assignments is not possible. Later assignments had algorithms with many more parameters than the earlier assignments, and an algorithm with more parameters would lead to more arguments per routine. It is important to note that despite varying number of arguments, the average number of arguments per routine remain low throughout the assignments.

\begin{figure}[h!]
	\centering
    \subfigure[Eiffel]
    {	
	\begin{tikzpicture}
	\begin{axis}[
 	 width=4.3cm, height=3.5cm,
	 scaled ticks=false, tick label style={/pgf/number format/fixed},
	 xbar, 
	 xmin = 1, xmax = 5,
	 ymin = 0, ymax = 100,
	 ytick = {0, 20, 40, 60, 80, 100},
	 yticklabels = {0\%, 20\%, 40\%, 60\%, 80\%, 100\%},
	 xtick = data,
	 xticklabel style = {inner sep=0pt,anchor=north east,rotate=45},
	 xticklabels = {A1, A2, A3, A4},
	 ybar interval=0.4,
	 bar width=2pt,
	 grid = major
	 ]
	  \addplot [color = red!70!black!70!white, fill = red!70!black!70!white]	
	    coordinates{(1,73)
	                (2,9)
	                (3,17)
	                (4,0)
	                (5,0)};
	 \end{axis}
	\end{tikzpicture}
	\label{subfig:parameiffel}
	}
	\subfigure[C++]
    {
	\begin{tikzpicture}
	\begin{axis}[
	width=3.9cm, height=3.5cm,
	 scaled ticks=false, tick label style={/pgf/number format/fixed},
	 xbar, 
	 xmin = 1, xmax = 4,
	 ymin = 0, ymax = 100,
	 ytick = {0, 20, 40, 60, 80, 100},
	 yticklabels = {0\%, 20\%, 40\%, 60\%, 80\%, 100\%},
	 xtick = data,
	 xticklabel style = {inner sep=0pt,anchor=north east,rotate=45},
	 xticklabels = {A2, A3, A4},
	 ybar interval=0.4,
	 bar width=2pt,
	 grid = major
	 ]
	  \addplot [color = black!50, fill = black!50]
	    coordinates{(1,36)
	                (2,50)
	                (3,33)
	                (4,0)};
	 \end{axis}
	\end{tikzpicture}
	\label{subfig:paramc++}
	}
	\vspace{-4mm}
	\caption{Lack of parametrization. N=11 for assignments 1 and 2. N=6 for assignments 3 and 4.}
	\vspace{-2mm}
	\label{fig:parametrization}
\end{figure}
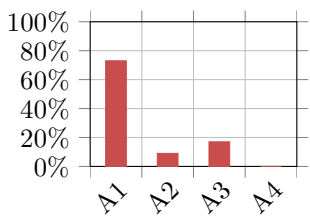
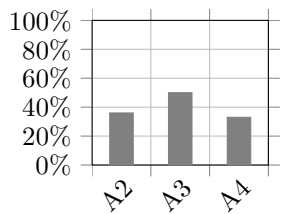

We also counted the number of submitted solutions that lacked parametrization. The results are shown in Fig.~\ref{fig:parametrization}. In the first assignment, over 70\% of the solutions lacked parametrization. After we pointed out the mistake during the feedback phase, the occurrence dropped significantly.

Hard-coded values and magic numbers were also prominent in the first assignment. During the feedback session, we recommended the students to introduce variables and use language support for \textit{constants}. Fig.~\ref{fig:hardcoded} shows how the percentage of routines with hard-coded values changed over the assignments. There is a significant drop from the first to the second assignment in Eiffel.

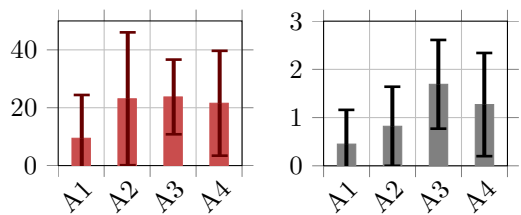
\begin{figure}[h!]
	\centering
    \subfigure[Total commits]
    {	
	\begin{tikzpicture}
	\begin{axis}[
	width=4cm, height=3.5cm,
	 scaled ticks=false, tick label style={/pgf/number format/fixed},
	 xbar, 
	 xmin = 1, xmax = 5,
	 ymin = 0, ymax = 50,
	 xtick = data,
	 xticklabel style = {inner sep=0pt,anchor=north east,rotate=45},
	 xticklabels = {A1, A2, A3, A4},
	 ybar interval=0.4,
	 bar width=2pt,
	 grid = major
	 ]
	  \addplot [color = red!70!black!70!white, fill = red!70!black!70!white]
	  [ error bars/.cd, y dir=both, y explicit,
	    error bar style={line width=1.5pt, red!40!black, xshift=3.1mm},
	    error mark options = {rotate = 90, line width=1pt, mark size = 3pt, red!40!black} ]	
	    coordinates{(1,9.45) +- (0,14.96)
	                (2,23.09) +- (0,22.96)
	                (3,23.73) +- (0,12.91)
	                (4,21.55) +- (0,18.12)
	                (5,0)};
	 \end{axis}
	\end{tikzpicture}
	\label{subfig:svn}
	}
	\subfigure[Daily commits]
    {
	\begin{tikzpicture}
	\begin{axis}[
	width=4cm, height=3.5cm,
	 scaled ticks=false, tick label style={/pgf/number format/fixed},
	 xbar, 
	 xmin = 1, xmax = 5,
	 ymin = 0, ymax = 3,
	 xtick = data,
	 xticklabel style = {inner sep=0pt,anchor=north east,rotate=45},
	 xticklabels = {A1, A2, A3, A4},
	 ybar interval=0.4,
	 bar width=2pt,
	 grid = major
	 ]
	  \addplot [color = black!50, fill = black!50]
	  [ error bars/.cd, y dir=both, y explicit,
  	    error bar style={line width=1.5pt, black, xshift=3mm},
	    error mark options = {rotate = 90, line width=1pt, mark size = 3pt, black} ]	
	    coordinates{(1,0.45) +- (0,0.71)
	                (2,0.82) +- (0,0.82)
	                (3,1.69) +- (0,0.92)
	                (4,1.27) +- (0,1.07)
	                (5,0)};
	 \end{axis}
	\end{tikzpicture}
	\label{subfig:svnday}
	}
	\vspace{-4mm}
	\caption{SVN repository usage.}
	\vspace{-2mm}
	\label{fig:svn}
\end{figure}

Fig.~\ref{fig:svn} shows the SVN repository usage. As half of the students had never used any configuration management tool, the initial SVN usage was relatively low. The students who were familiar with SVN committed their incremental changes while those who were new made commits sporadically or not at all. After the first assignment, however, the students started using SVN more regularly. The number of total commits per assignment jumped after the first assignment and remained steady. The daily usage of SVN, the number of commits divided by the number of days, increases over the first three assignment before dropping for the fourth assignment. The drop may have been caused by time crunch at the end of the semester.

\section{Discussion}
\label{sec:discussion}

This section discusses meaning of the results in terms of the students' knowledge gain in software engineering.

\subsection{Pre-class survey}
The pre-survey revealed that majority of the students came to the class with limited experience in programming and little to no exposure to key software engineering tools. The discrepancy in software engineering knowledge and experience between computer science students and non-computer science students was quite stark. Despite the importance of software engineering in other fields, some non-computer science students had only programmed in class and had no exposure to object-oriented programming. Given that the course is multidisciplinary and hence requires knowledge in many different fields, educational background and programming experience of a student had little effect on the student's final performance in the class.

\subsection{Post-class survey}
We analyzed two open-ended questions of the post-survey. Responses to the first question revealed that the students used several concepts and tools during the semester. Most students mentioned concepts related to software quality, including reusability, abstraction, and documentation. This shows that our students developed some understanding of software quality, one of our main objectives. Software tools with editor and debugger as sub-themes was also frequent. As the question contained the word \emph{tools}, identifying tools as a theme is not surprising; however, given that we devoted a lecture on software engineering tools, we expected our students to have used more than editor and debugger. The third theme, programming paradigm, with its two sub-themes, object-oriented programming and concurrency, show that the students became more aware of these paradigms compared to the beginning of the course.

Our analysis of the second question revealed that most students improved software development and/or project management. In software development, the students indicated that they now think about software architecture and quality and develop software incrementally. This result was as we had hoped. What we did not expect as much was their improvement in project management skills. Many indicated that they now manage their time better and set priority when working. This is a positive by-product of the students working on challenging assignments with limited time, and these skills are definitely useful in software engineering.

In general, we faced several issues in analyzing the responses. One major issue is that some students gave long, detailed answers to the survey questions whereas others did not. Short responses contain, inevitably, limited information. While it is likely that the items mentioned in the response are those that the students found most important, it is difficult to judge if the response captures the complete set. Other issue is that the questions were narrowly formulated. While this reduces potential misunderstanding, it also limits variety of the responses. This was especially true for the second question, to which most students gave yes or no answer with a short explanation. Last, the post-survey contained many other questions not reported in this paper, and the sheer number of question may have caused some students to provide only a brief response to each question.

\subsection{Assignments}
Quantitative analysis of the assignments using software quality metrics revealed that the students improved after the first assignment in some areas and remained the same in others. Among the five metrics we used for the analysis, percentage of comments and number of arguments per routine did not reveal much in terms of improvement of the students' software engineering techniques. Parametrization, hard-code variables, and SVN usage showed that the students made a leap after the first assignment. Likely cause of improvement is the individual feedback session. After the first feedback session, many students remarked that no one had given them the kind of tips we gave. In the subsequent assignments, quality of student-generated software depended more on difficulty of the assignments and students' willingness to put their knowledge into practice. When we pointed out mistakes during the assignment 2 and 3 feedback sessions, the students often showed their awareness of the mistakes and blamed the time-constraint for not applying their knowledge into practice. 

The analysis revealed that in general, the code written in Eiffel were of higher quality than those in C++. This may be because for each assignment, C++ portion required the students to implement a new robotics algorithm whereas Eiffel portion required them to extend the previous version. We also observed high variability among the submitted software. Some students submitted meticulously cleaned-up, refactored, and commented code while others submitted code with commented-out code blocks and To-Do notes. Despite the variability, most students improved after the first feedback session.

\subsection{General remark}
Relating the survey results and the analysis of the assignments shows that our students came in with limited knowledge of software engineering and left with an understanding of software engineering principles and techniques. The course focused on modularity and documentation as the main indicators of quality software, and the survey responses revealed that the students' appreciation of software quality is directly related to these. Minimization of hard-coded variables and introduction of parametrization in their assignment reflect that the students tried to put their knowledge into practice. The students also claimed that they had changed to incremental development, and increase in the frequency of SVN commits is in line with their statement. Overall, the students may not have internalized all the pitfalls, but they have gained an understanding of proper software engineering and applied some into practice.

\section{Threats to Validity}
\label{sec:threats}

There are several limitations and threats to validity of this study. An obvious limitation of the study is that data are drawn from a single course offering at one university. While the pilot study provides some understanding of the effect of teaching software engineering in a robotics programming course, longer and broader study is necessary to generalize the claim. In addition, the study's small data size limits generalization of the results.

The study contains several potential sources of bias. First, as an elective course, no student was required to take the course. The students may have been more motivated to learn about software engineering in robotics setting than regular students, and this may bias our results towards better learning outcome. Second, the authors designed, implemented, and executed the course described in the paper. Even though we tried to be neutral in analyzing data, the results may nonetheless be biased towards success. Lastly, as the students got to know the authors very well by the end of the course, their post-survey responses may have been influenced by their feeling towards the authors. In fact, top third of the students gave longer and more-detailed responses to the survey than most other students, resulting in their opinion being more reflected in the qualitative analysis. This may bias towards better post-survey results.

In our analysis, we assume that the assignments capture their understanding of software engineering. In reality, the correlation between the submitted code and students' knowledge is not clear. Many students found the course challenging and time-consuming and tried to optimize the way they spend their time. Consequently, software they submitted is not necessarily representative of what they actually learned in class. Due to time limitation, students may have not applied everything they have learned but instead only those that result in better grade for the effort. In addition, the last two assignments were completed in a team of two students, and it is not clear how the knowledge of two students maps to a single piece of software. 

\section{Conclusions}
\label{sec:conclusions}

This paper reported and evaluated a newly-developed robotics programming course that emphasizes proper software engineering -- modularity and documentation -- in robotics context. The course covers topics of software engineering and robotics equally, and the students learn to apply proper software engineering techniques by implementing four different robotics algorithms for an educational robot. To understand the effect of the course, we conducted a pre-survey and a post-survey and analyzed the four assignments using software quality metrics. The analysis of survey responses and student-generated code showed that the students gained some understanding of software engineering principles and techniques.

Based on our analysis in this pilot study, we believe that students \emph{can} learn software engineering by programming a robot, and software engineering education is important to students outside of computer science. Fully understanding the effect of teaching software engineering in a robotics setting requires more extensive research. To validate the results of this pilot study, we plan to continue our evaluation of the course in the subsequent offerings and collect data over a longer period. It would be especially interesting to understand the long-term effect of our course to computer science students and other students alike.

\section{Acknowledgments}
The research leading to these results has received funding from the European Re-
search Council under the European Union's Seventh Framework Programme
(FP7/2007-2013) / ERC Grant agreement no. 291389, Hasler Foundation under SmartWorld initiative / Roboscoop: concurrent robotics framework project, and ETH Zurich under Innovedum / project no. 733.

\bibliographystyle{abbrv}
\bibliography{bibliography}
\end{document}